\documentclass[a4paper]{article}

\usepackage[margin=3cm]{geometry}
\usepackage[numbers]{dmnatbib}

\usepackage{amsthm}
\theoremstyle{theorem}
\newtheorem{theorem}{Theorem}
\newtheorem{lemma}[theorem]{Lemma}
\newtheorem{corollary}[theorem]{Corollary}
\theoremstyle{definition}
\newtheorem{definition}[theorem]{Definition}

\newenvironment{pf*}{\begin{proof}}{\end{proof}}

\usepackage{listings}
\usepackage{graphicx}
\usepackage{amssymb}
\usepackage[mathscr]{euscript}
\usepackage{subfigure}
\usepackage{multirow}
\usepackage{array}

\usepackage{tikz}
\usetikzlibrary{shapes}

\tikzstyle{block} = [rectangle, inner sep=1ex, draw]
\tikzset{axis/.style={->}}
\tikzset{boundary/.style={draw}}
\tikzset{widen/.style={boundary, black, fill=red!70, fill opacity=0.8}}
\tikzset{entry/.style={boundary, black, fill=blue!50, fill opacity=0.6}}
\tikzset{newpoly/.style={boundary, fill=pink, fill opacity=0.8}}
\tikzset{concrete/.style={draw, ultra thick, dashed}}

\newcommand{\vertex}[1]{
	\node[coordinate] (temp)at #1 {};
	\fill (temp) circle(0.2);
}

\newcommand{\myframe}{
    \def\xmax{5}
    \def\ymax{10}
    \draw[help lines] (\xmax, -1) -- (\xmax, \ymax+1);
    \draw[help lines] (-1, \ymax) -- (\xmax+1, \ymax);
    \draw[axis] (-1,0) -- (\xmax+1.5,0) node(xline)[above] {$i$};
    \draw[axis] (0,-1) -- (0,\ymax+2.5) node(yline)[right] {$j$};
    \clip (-1,-1) rectangle (\xmax+1,\ymax+1);
}

\usepackage{hyperref}

\newcommand{\widening}{\triangledown}
\newcommand{\narrowing}{\bigtriangleup}
\newcommand{\calS}{\mathcal{S}}
\newcommand{\calV}{\mathcal{V}}
\newcommand{\abstr}[1]{#1^\sharp}
\newcommand{\lfp}{\textrm{lfp~}}
\newcommand{\tfun}{\Psi}
\newcommand{\parts}[1]{\mathscr{P}(#1)}
\newcommand{\ZZ}{\mathbb{Z}}
\newcommand{\QQ}{\mathbb{Q}}

\title{Stratified Static Analysis Based on Variable Dependencies\thanks{This work was partially supported by ANR project ``\href{http://asopt.inrialpes.fr/}{ASOPT}}}

\author{\href{http://www-verimag.imag.fr/~monniaux/}{David Monniaux}\thanks{CNRS / VERIMAG; \href{http://www-verimag.imag.fr/}{VERIMAG} is a joint laboratory of \href{http://www.cnrs.fr/}{CNRS} and \href{http://www.ujf-grenoble.fr/}{Universit\'e Joseph Fourier}} \and \href{http://www.jleguen.info/}{Julien Le Guen}\thanks{VERIMAG \& STMicroelectronics}}

\begin{document}
\maketitle

\begin{abstract}
In static analysis by abstract interpretation, one often uses \emph{widening operators} in order to enforce convergence within finite time to an inductive invariant. Certain widening operators, including the classical one over finite polyhedra, exhibit an unintuitive behavior: analyzing the program over a subset of its variables may lead a more precise result than analyzing the original program!
In this article, we present simple workarounds for such behavior.
\end{abstract}

\section{Introduction} 
During experiments, we found examples over which classical polyhedral analysis \cite{CousotHalbwachs78}, even with alternative widenings \cite{BagnaraHRZ05SCP}, would fail to discover some simple program invariants, which could sometimes even be discovered by interval analysis. This would even happen on simple loops, e.g. \lstinline|for(int i=0; i<N; i++)|, if the loop contained a nested loop not touching \lstinline|i|: the analysis would not discover $\lstinline|i|\geq 0$! It is counter-intuitive that difficulties in analyzing the behavior of the program on other variables should lead to imprecise results for~\lstinline|i|.

In some of these examples, such as this simple loop, the lost invariants could be easily recovered by syntactic pattern-matching, but such techniques are brittle. We therefore searched for techniques inspired by our intuition that poor results on certain variables should not impact variables not depending on them.

\subsection{Generalities and Notations}
We consider the strongest invariant of a loop (or, more generally, of a program), defined as the least fixed point $\lfp \tfun$ of a monotone operator~$\tfun$ over sets of program states~\cite{CousotCousot_JLC92}. For instance, in program~\ref{lst:loop42squared}, the strongest invariant of the loop is the least fixed point in $(\parts{\ZZ \times \ZZ}, \subseteq)$ of the operator
\begin{equation}
\tfun(X) = \{ (1, 0) \} \cup \{ (i+1, j+i) \mid (i, j) \in X \land i \leq 5 \}
\end{equation}

Explicit-state model-checking computes such invariants as explicitly represented sets of states (that is, for each state there exists some little data structure). Implicit-state model checking uses compact representations of such sets, such as binary decision diagrams, and computes the least solution of $\tfun(X)=X$ by finding the limit of the ascending sequence $X_0 = \emptyset$, $X_{n+1}=\tfun(X_n)$; for systems with at most $n$ states, this limit is reached within at most $n$ iterations. For infinite state systems such as software programs%
\footnote{One of the authors once heard the remark that a program without dynamic allocation or recursion was just a finite-state automaton, thus all properties are decidable, including halting. For the purpose of practical analysis, except for very small and simple programs, such state spaces are so large that they should be treated as infinite.}
such an approach is infeasible, because (a) the sets of states $X_i$ may be large (or even infinite, if infinite nondeterminism is used) (b) the sequence may not converge within a finite number of iterations.

Abstract interpretation \cite{CousotCousot_JLC92,DBLP:journals/cl/CortesiZ11} solves point (a) by replacing arbitrary sets of states by \emph{over-approximations}; for instance, a set of points in $\ZZ^n$ or $\QQ^n$ may be replaced by an enclosing convex polyhedron \cite{Halbwachs_PhD,CousotHalbwachs78,Polka:FMSD:97}. A given analysis thus restricts itself to a given \emph{abstract domain} of sets of states; in this article, we focus, as an example, on the domain of polyhedra, but there exist many other abstract domains, for numerical \cite{Mine_PhD} or non-numerical states. The operator $\tfun$ on concrete states is replaced by an abstract operator $\abstr{\tfun}$, satisfying a soundness condition $\tfun(\abstr{X}) \subseteq \abstr{\tfun}(\abstr{X})$ for all~$\abstr{X}$.%
\footnote{Some presentations of abstract interpretation distinguish the abstract element $\abstr{X}$ from the set of states $\gamma(\abstr{X})$ that it represents. In this article, we chose not to, in order to simplify notations.}

Problem (b), that is, failure for the sequence $\abstr{X}_{n+1}=\abstr{\tfun}(\abstr{X}_n)$ to become stationary, remains if the abstract domains contains infinite
strictly ascending sequences;%
\footnote{Again, for practical purposes, it suffices that there exist exceedingly long finite ascending sequences for analysis to become unfeasible.} this is for instance the case of the domain of convex polyhedra. Some form of convergence acceleration is thus needed.
Starting with $\abstr{u}_0 = \emptyset$, \emph{upwards iterations with widening} \cite{DBLP:journals/cl/CortesiZ11,CousotCousot_JLC92} compute%
\footnote{Following the usage in APRON \cite{DBLP:conf/cav/JeannetM09}, our definition of $u \widening v$ assumes that $u \subseteq v$; if this is not the case, use $u \widening (u \sqcup v)$ instead.}
\begin{equation}\label{eqn:iterations}
\abstr{u}_{n+1} = \abstr{u}_n \widening (\abstr{u}_n \sqcup \abstr{\tfun}(\abstr{u}_n))
\end{equation}

$x \sqcup y$ is such that $x, y \subseteq x \sqcup y$ (in the case of polyhedra, $\sqcup$ is generally taken to be the convex hull), and $\widening$ is a \emph{widening operator}, such that for all $x \subseteq y$, $y \subseteq x \widening y$ (\emph{soundness} property), and any sequence of the form $\abstr{u}_{n+1} = \abstr{u}_n \widening \abstr{v}_n$, where $\abstr{v}_n$ is any other sequence, is stationary: after a certain $N$, it is constant (\emph{termination} property). Then, $\tfun(\abstr{u}_N) \subseteq \abstr{\tfun}(\abstr{u}_N) \subseteq \abstr{u}_N \widening (\abstr{u}_N \sqcup \abstr{\tfun}(\abstr{u}_N)) = \abstr{u}_N$, thus $\tfun(\abstr{u}_N) \subseteq \abstr{u}_N$, which means that $\abstr{u}_N$ is an inductive invariant of the program, in which the strongest invariant is included.

Once an inductive invariant $\abstr{u}_N$ is obtained, it may be refined by \emph{narrowing} iterations, which in practice generally consist in computing ${\abstr{\tfun}}^k(\abstr{u}_N)$ until the sequence becomes stationary or $k$ exceeds a preset limit.

Widening operators have various unpleasant properties. The best known is that they bring \emph{imprecision}: the result of widening/narrowing iterations may be strictly larger than the least element of the abstract domain that is an inductive invariant, let alone an invariant (in Sec.~\ref{sec:related_work} we shall list some alternative approaches that do not suffer from this inconvenience, at the expense of generality). The contribution of this article is a generic method to reduce some of the imprecision induced by widening.

\subsection{Motivating Example} 
\label{part:motivation_example}

Classical polyhedral analysis \cite{CousotHalbwachs78},%
\footnote{One may try examples on B.~Jeannet's online Interproc analyzer at \url{http://pop-art.inrialpes.fr/interproc/interprocweb.cgi}}
when applied to Listing~\ref{lst:loop42}, discovers that $i \geq 1 \land i \leq 5$ is an invariant at the head of the loop. 
Yet, running the same analysis on Listing~\ref{lst:loop42squared} yields $i \leq 5$ but not $i \geq 1$.

\noindent
\begin{minipage}[b]{0.45\textwidth}
\begin{lstlisting}[label=lst:loop42,caption={Loop until 5}]
int i=1;
while (i<=5) {
  i=i+1;
}
\end{lstlisting}
\begin{lstlisting}[label=lst:loop42squared,caption={$j=i(i+1)/2$}]
int i=1, j=0;
while (i<=5) {
  j=j+i;
  i=i+1;
}
\end{lstlisting}
\end{minipage}\hfill
\begin{minipage}[b]{0.45\textwidth}
    \centering
    
    \begin{tikzpicture}
        \tikzset{ppoint/.style={
            shape=circle, 
            draw, 
            ultra thick, 
            node distance=4em,
            font=\footnotesize
        }}

        \node[ppoint, label=left:\footnotesize$n0$] (n0) {};
        \node[below of=n0, ppoint, label=left:\footnotesize$n1$] (n1) {};
        \node[below right of=n1, ppoint, label=left:\footnotesize$n2$] (n2) {};
        \node[below of=n2, ppoint, label=left:\footnotesize$n3$] (n3) {};
        \node[below left of=n1, ppoint, label=left:\footnotesize$n4$] (n4) {};

        \draw[->] (n0) -- (n1)
            node[right,midway,text width=4em] {
                \footnotesize
                $i \leftarrow 1$
                $j \leftarrow 0$
            };
        \draw[->] (n1) -- (n2)
            node[right,midway] {
                \footnotesize
                $i \leq 5$
            };
        \draw[->] (n2) -- (n3)
            node[right,midway,text width=5em] {
                \footnotesize
                $j \leftarrow j + i$
                $i \leftarrow i + 1$
            };
        \draw[->] (n3.east) .. controls +(right:8em) and +(right:10em) ..  (n1.east);
        \draw[->] (n1) -- (n4)
            node[left,midway] {
                \footnotesize
                $i > 5$
            };

    \end{tikzpicture}

    \label{fig:program}
\end{minipage}

This example is not fortuitous: it models how to address consecutive lines of a matrix in lower triangular packed storage mode. In that memory-effective approach, the matrix is stored in memory as a unidimensional array, each line next to the preceding one, and line number \lstinline|i| only uses \lstinline|i| positions in the array: \lstinline|j| is the index of the start of the line in the array.

Program~\ref{lst:loop42} is an abstraction of Program~\ref{lst:loop42squared}: each execution of the latter maps to an execution of the former. Yet, the analysis of the former produces a more precise loop invariant than the analysis of the latter. This is an example of the non-monotonicity of analyzes using widenings, a long-known phenomenon~\cite[ex.~11]{DBLP:conf/plilp/CousotC92}: a more precise abstraction may ultimately lead to less precision in the final analysis result.

Analysis of Program~\ref{lst:loop42squared} with the basic upwards iteration and widening scheme (widening at every iteration) \cite{CousotCousot_JLC92}, using the standard widening on polyhedra,%
\footnote{The standard widening on polyhedra $P_1 \widening_S P_2$, in intuitive terms, suppresses from $P_2$ constraints not present in $P_1$. In reality, its correct definition contains subtleties regarding polyhedra of dimension less than the dimension of the space, and the original definition \cite{CousotHalbwachs78} had to be corrected \cite{Halbwachs_PhD}. \cite{BagnaraHRZ05SCP} recalls the corrected definition.}
yields the successive polyhedra
\begin{itemize}
\item $i=1 \land j=0$
\item $-i+j \geq -1 \land i \geq 1$: draw a line through the first two reachable states and obtain a polyhedron in $(i,j)$ generated by vertex $(1,0)$ and ray $(1,1)$;
\item $-i+j \geq -1 \land 7i-4j \geq 7$: polyhedron in $(i,j)$ generated by vertex $(1,0)$ and rays $(1,1)$ and $(4,7)$.
\end{itemize}
So far, so good: such polyhedra still imply $i \geq 1$. At the next iteration, however, this constraint is lost and one gets the polyhedron $-i+j \geq -1$, and finally $\top$, the whole plane. The constraint $i \leq 5$ is recovered by one step of downwards iteration. Analysis with the improved widening proposed by Bagnara et al. \cite{BagnaraHRZ05SCP}, as implemented in the Parma Polyhedra Library, yields a different iteration sequence, but still reaches $\top$ at the end.

If one runs a polyhedral analysis on Program~\ref{lst:loop42}, one gets the inductive invariant $1 \leq i \leq 5$, which is also valid for Program~\ref{lst:loop42squared}. Intersecting this invariant with the output of the widening in the analysis of  Program~\ref{lst:loop42squared} yields a reasonably precise polyhedron (Table~\ref{tab:table1}).

%

\begin{table}[t]\footnotesize
    \tikzset{every picture/.style={scale=0.1}}
    \centering
        %
%

    \begin{tabular}{|c|m{2.6cm}|*{6}{m{1cm}|}}
        \hline
        Analysis & Node 
            & $\abstr{u}_0$ & $\abstr{u}_1$ & $\abstr{u}_2$ 
            & $\abstr{u}_3$ & $\abstr{u}_4$ & $\abstr{u}_5$\\ \hline
        \multirow{2}{*}{Classic} & 
            $n1$ (entry) & 
                \begin{tikzpicture}
                    \myframe
                    \vertex{(1,0)}
                \end{tikzpicture}
                & 
                \begin{tikzpicture}
                    \myframe
                    \vertex{(1,0)}
                    \vertex{(2,1)}
                    \path[entry] (1,0) -- (2,1);
                \end{tikzpicture}
                & 
                \begin{tikzpicture}
                    \myframe
                    \vertex{(1,0)}
                    \vertex{(2,1)}
                    \vertex{(5,7)}
                    \path[entry] (1,0) -- (2,1) -- (5,7) -- cycle;
                \end{tikzpicture}
                & 
                \begin{tikzpicture}
                    \myframe
                    \vertex{(1,0)}
                    \vertex{(2,1)}
                    \vertex{(5,7)}
                    \vertex{(5,37/4)}
                    \path[entry] (1,0) -- (2,1) -- (5,7) -- (5, 37/4) -- cycle;
                \end{tikzpicture}
                & 
                \begin{tikzpicture}
                    \myframe
                    \vertex{(5,7)}
                    \path[entry] (-2,-7) -- (5,7) -- (5, 15) -- (-5, 15);
                \end{tikzpicture}
                & 
                \begin{tikzpicture}
                    \myframe
                    \vertex{(5,0)}
                    \path[entry] (-2,-2) -- (5,-2) -- (5,0) -- (5, 15) -- (-2, 15);
                \end{tikzpicture}
                \\ \cline{2-8}
            & $n1$ (after $\widening$ or $\narrowing$)
                & 
                \begin{tikzpicture}
                    \myframe
                    \vertex{(1,0)}
                \end{tikzpicture}
                & 
                \begin{tikzpicture}
                    \myframe
                    \vertex{(1,0)}
                    \path[newpoly] (1,0) -- (10,9);
                \end{tikzpicture}
                & 
                \begin{tikzpicture}
                    \myframe
                    \vertex{(1,0)}
                    \path[newpoly] (1,0) -- (10,9) -- (8,14) -- cycle;
                \end{tikzpicture}
                & 
                \begin{tikzpicture}
                    \myframe
                    \vertex{(1,0)}
                    \path[newpoly] (-5,-6) -- (1,0) -- (8,7) -- (8, 15) -- (-5, 15);
                \end{tikzpicture}
                & 
                \begin{tikzpicture}
                    \myframe
                    \path[newpoly] (-2,-2) rectangle (15,15);
                \end{tikzpicture}
                & 
                \begin{tikzpicture}
                    \myframe
                    \vertex{(5,0)}
                    \path[newpoly] (-2,-2) -- (5,-2) -- (5,0) -- (5, 15) -- (-2, 15);
                \end{tikzpicture}
                \\ \hline
        \hline
        \multirow{2}{*}{Stratified} 
            & $n1$ (entry)
                & 
                \begin{tikzpicture}
                    \myframe
                    \vertex{(1,0)}
                \end{tikzpicture}
                & 
                \begin{tikzpicture}
                    \myframe
                    \vertex{(1,0)}
                    \vertex{(2,1)}
                    \path[entry] (1,0) -- (2,1);
                \end{tikzpicture}
                & 
                \begin{tikzpicture}
                    \myframe
                    \vertex{(1,0)}
                    \vertex{(2,1)}
                    \vertex{(5,7)}
                    \path[entry] (1,0) -- (2,1) -- (5,7) -- cycle;
                \end{tikzpicture}
                & 
                \begin{tikzpicture}
                    \myframe
                    \vertex{(1,0)}
                    \vertex{(2,1)}
                    \vertex{(5,7)}
                    \vertex{(5,37/4)}
                    \path[entry] (1,0) -- (2,1) -- (5,7) -- (5, 37/4) -- cycle;
                \end{tikzpicture}
                & 
                \begin{tikzpicture}
                    \myframe
                    \vertex{(1,0)}
                    \vertex{(2,1)}
                    \vertex{(5,7)}
                    \path[entry] (1,15) -- (1,0) -- (2,1) -- (5,7) -- (5,15);
                \end{tikzpicture}
                & 
                \begin{tikzpicture}
                    \myframe
                    \vertex{(1,0)}
                    \vertex{(2,1)}
                    \vertex{(3,3)}
                    \vertex{(5,9)}
                    \path[entry] (1,15) -- (1,0) -- (2,1) -- (3, 3) -- (5, 9) -- (5,15);
                \end{tikzpicture}
                \\ \cline{2-8}
            & $n1$ (after $\widening$ or $\narrowing$) and intersection with previous stratum
                & 
                \begin{tikzpicture}
                    \myframe
                    \vertex{(1,0)}
                \end{tikzpicture}
                & 
                \begin{tikzpicture}
                    \myframe
                    \vertex{(1,0)}
                    \vertex{(5,4)}
                    \path[newpoly] (1,0) -- (5,4);
                \end{tikzpicture}
                & 
                \begin{tikzpicture}
                    \myframe
                    \vertex{(1,0)}
                    \vertex{(5,4)}
                    \vertex{(5,7)}
                    \path[newpoly] (1,0) -- (5,4) -- (5,7) -- cycle;
                \end{tikzpicture}
                & 
                \begin{tikzpicture}
                    \myframe
                    \vertex{(1,0)}
                    \vertex{(5,4)}
                    \path[newpoly] (1,15) -- (1,0) -- (5,4) -- (5, 15);
                \end{tikzpicture}
                & 
                \begin{tikzpicture}
                    \myframe
                    \vertex{(1,0)}
                    \vertex{(2,1)}
                    \vertex{(5,7)}
                    \path[newpoly] (1,15) -- (1,0) -- (2,1) -- (5,7) -- (5,15);
                \end{tikzpicture}
                & 
                \begin{tikzpicture}
                    \myframe
                    \vertex{(1,0)}
                    \vertex{(2,1)}
                    \vertex{(3,3)}
                    \vertex{(5,9)}
                    \path[newpoly] (1,15) -- (1,0) -- (2,1) -- (3, 3) -- (5, 9) -- (5,15);
                \end{tikzpicture}
            \\ \hline
    \end{tabular}
        \caption{Comparison of classic static analysis (upward iterations with
        widening $\widening$ followed by descending iterations) and stratified static
        analysis on Program~\ref{lst:loop42squared}. Classic analysis loses the
        constraint $i \geq 1$ and finds $\top$ in 5 iterations. The upper bound
        $i \leq 5$ is found with one narrowing iteration. Stratified analysis
        on the stratum consisting of variable $i$ first finds $1 \leq i \leq
        5$. Then, it analyzes stratum {$i$, $j$} and intersect with result of
        stratum {$i$}. A fixed point is found after 4 iterations
        ($\abstr{u}_3$, last line). The table also shows the polyhedra found
        after two narrowing iterations. The resulting polyhedron, even without
        narrowing iterations, is much more precise than the one found by
        classic analysis.}
    \label{tab:table1}
\end{table}

Thus, the basic idea of our method: \emph{run preliminary analyzes over abstractions of the program obtained by removing some of the variables}, in order to refine the analysis of the complete program.
In order to further convey our intuition, let us remark that Prog.~\ref{lst:loop42squared} is the result of \emph{loop fusion} over the following program~:
\begin{lstlisting}
for(i=1; i<=5; i++) t[i]=i;
for(i=1; i<=5; i++) j += t[i];
\end{lstlisting}
Normal forward polyhedral analysis on this program will find good invariants for both loops. In particular, the second loop may not perturb analysis of the first loop. It seems reasonable that the same applies to the code after loop fusion.

The same code could have been the result of the compilation into C of a data-flow program (e.g. Simulink or Lustre) consisting in a ramp generator and an integrator:
\begin{center}
\begin{tikzpicture}[node distance=3cm]
\node[block] (ramp) {ramp $1\dots 5$} ;
\node[block, right of=ramp] (integrator) {$\Sigma$} ;
\path (ramp) edge[->] (integrator);
\end{tikzpicture}
\end{center}
Again, it seems natural that the analysis of the integrator should not hamper the analysis of the ramp.

\section{Stratified Analysis} 
\label{sec:stratified_analysis}
We have investigated two approaches. In \emph{stratified analysis}, we successively perform several static analyzes by abstract interpretation, the results from each analysis being used to refine the following ones. In \emph{stratified widening}, a single analysis pass is performed, but with a widening improving on and derived from the traditional widening on polyhedra.

\subsection{Dependency Strata}
\label{part:strata}
We consider a set $\calS$ of subsets of the set of variables $\calV$ of the program, such that $\calV \in \calS$; we order it by inclusion.
An \emph{immediate predecessor} of $S \in \calS$, denoted by $S' \prec S$, is $S'$ such that $S' \subsetneq S$ and there is no $S''$ such that $S' \subsetneq S'' \subsetneq S$.

In practice, if we have a relationship $v_1 \rightarrow v_2$ meaning ``$v_1$ flows into $v_2$ through some computation'' or ``$v_2$ depends on $v_1$'', then the elements of $\calS$ are, in addition to $\calV$ itself, subsets $S$ of $\calV$ closed by: if $v \in S$ and $v' \rightarrow v$, then $v' \in S$. One way to construct such subsets is to compute for each variable $v$ the set $S(v) = \{ v' \mid v' \rightarrow v \}$, and add this set to $\calS$ unless it is already present. For better efficiency, one computes the strongly connected components of $\rightarrow$, and takes $S(v)$ for one $v$ in each component.

Note that $\rightarrow$ needs not be the semantics dependency relation, which takes into account both data and control dependencies. In intuitive (and imprecise) terms, a variable $x$ is said to be data-dependent on a variable $y$ if $x$ is assigned to by an expression where $y$ appears; a variable $x$ is said to be control-dependent on a variable $y$ if $x$ is assigned in a program branch executed or not executed according to the value of~$y$. Collecting all program elements on which a variable depends, through data or control dependencies, is known as \emph{slicing}~\cite{DBLP:journals/tse/Weiser84}. If $\rightarrow$ takes into account all dependencies, then $S(v)$ is the slice of variables on which $v$ depends.

A helpful intuition of our method is that it performs analyzes on program slices of increasing size; but this is somewhat misleading, because we do not make any assumption on $\rightarrow$ and thus it does not necessarily reflect all dependencies. In particular, ignoring control dependencies, compared conventional slicing, may produce simpler slices, of a more manageable size --- X.~Rival, when developing the Astr\'ee static analyzer, observed that, for many variables, the slice corresponded to approximately 80\% of the code, thus slicing did not significantly simplify the program~\cite{DBLP:conf/sas/Rival05}.

\subsection{Informal Definition}
Let $S$ be a subset of the variables in program~$P$. We note $P_{| S}$ the program $P$ where all references to variables outside $S$ have been replaced by \lstinline|nondet()| nondeterministic choices.

\noindent\begin{minipage}{0.45\textwidth}
\begin{lstlisting}[title={Program $P$}]
int i=1, j=0;
while (i<=5) {
  j=j+i;
  if (j % 2 == 0) i=i+1;
}
\end{lstlisting}
\end{minipage}
\hfill
\begin{minipage}{0.45\textwidth}
\begin{lstlisting}[title={$P_{| S}$ for $S = \{ \lstinline|i| \}$}]
int i=1;
while (i<=5) {
  if (nondet()) i=i+1;
}
\end{lstlisting}
\end{minipage}

For any program $P$, let $C(P)$ be its collecting semantics: the set of reachable states of $P$. In order to simplify notations, for $S \subseteq S'$, we identify sets of states referring to the variables in $S$ with their completion by all values for variables in $S' \setminus S$. For any $S$, $P_{|S}$ is a safe abstraction of $P$: $C(P) \subseteq C(P_{|S})$. More generally, if $S \subseteq S'$, $C(P_{|S'}) \subseteq C(P_{|S})$.

For any program $P$, let $A(P)$ be the result of static analysis of $P$.
Correctness of the analysis means $C(P) \subseteq A(P)$.
Let $A(P,K)$ be the result of the static analysis of $P$ where the semantics of $P$ is restricted to states in $K$: in other words, all states outside of $K$ are removed from the transition relation. For any $K \supseteq C(P)$, $C(P) \subseteq A(P,K)$.

For each $S \in \calS$, we compute the intermediate analysis result $R(S)$ after all $R(S')$, $S' \prec S$, have been computed, as follows:
\begin{equation}\label{eqn:restricted_analysis}
R(S) = A\left(P_{|S},\bigcap_{S' \prec S} R(S')\right)
\end{equation}
Remark that in this formula, we could have made $S'$ to range over all predecessors without changing the result; however, this would have been less efficient.

By induction on the length of the $\prec$-chains, for all $S$, $R(S) \supseteq C(P_{|S})$. At the end, $R(\calV) \supseteq C(P)$ is a correct analysis result for the whole program; in fact, any $R(S) \supseteq C(P)$, so one can stop the analysis at any step, for instance because of a time limit.

This is the analysis performed in \S\ref{part:motivation_example}, with $\calS = \{ \{ \lstinline|i| \}, \{ \lstinline|i|, \lstinline|j| \} \}$.

\subsection{Formal Definitions and Variants}
\label{sec:formal}
Let $S \in \calS$. We assume that the result $R(S')$ of the analysis for all $S' \prec S$ has already been computed. Let $\abstr{K}=\bigcap_{S' \prec S} R(S')$; we assume that $\lfp \tfun \subseteq R(S')$ for all $S' \prec S$ and thus that $\lfp \tfun \subseteq \abstr{K}$.

The analysis described at Eqn.~\ref{eqn:restricted_analysis} is defined by the sequence:
\begin{equation}\label{eqn:restricted_iterations}
\abstr{u}_{n+1} = \abstr{u}_n \widening (\abstr{u}_n \sqcup (\abstr{\tfun}(\abstr{u}_n \cap \abstr{K}) \cap \abstr{K}))
\end{equation}
We compute the limit $R(S) = \abstr{u}_N$ of that stationary sequence, and output $\abstr{u}_N \cap \abstr{K}$.

Let us note $\tfun_{|A}(X) = \tfun(X \cap A) \cap A$. In other words, $\tfun_{|A}$ is $\tfun$ with everything outside of $A$ being discarded. The following lemma means that we do not change the strongest invariant by throwing out unreachable states in the definition of the semantics, which is intuitive.

\begin{lemma}\label{lem:restriction}
$\lfp \tfun = \lfp \tfun_{|A}$ for any $A \supseteq \lfp \tfun$.
\end{lemma}

\begin{proof}
$\lfp \tfun_{|A}$ is the limit of the ascending sequence defined by $X_0 = \emptyset$, $X_{n+1} =  \tfun_{|A}(X_n)$, $\lfp \tfun$ that of $Y_0 = \emptyset$, $Y_{n+1} =  \tfun(Y_n)$. By induction, for all $n$, $X_n = Y_n$.
\end{proof}

\begin{corollary}
$\abstr{u}_N$, and thus $\abstr{u}_N \cap \abstr{K}$, includes $\lfp \tfun$, that is, the reachable states.
\end{corollary}

\begin{pf*}{Proof}
Because $y \subseteq x \widening y$ and $y \subseteq x \sqcup y$ for all $x,y$,  $\abstr{\tfun}(\abstr{u}_N \cap \abstr{K}) \cap \abstr{K} \subseteq \abstr{u}_N$ and thus $\tfun_{\abstr{K}}(\abstr{u}_N) = \tfun(\abstr{u}_N \cap \abstr{K}) \cap \abstr{K} \subseteq \abstr{u}_N$. Thus, $\lfp \tfun_{\abstr{K}} \subseteq \abstr{u}_N$. The result follows from the lemma.
\end{pf*}

We conclude that, by induction over $\prec$, for all $S$, $\lfp \tfun \subseteq R(S)$. 

We shall now describe a subtly different iteration scheme, which supposes some additional properties of~$\widening$:

\begin{definition}\label{def:upto-termination}
 We say that $\widening$ satisfies the \emph{``up to'' termination condition} if for any fixed $\abstr{K}$, any $\abstr{u}_0 \subseteq \abstr{K}$, any sequence $\abstr{v}_n \subseteq \abstr{K}$ the sequence defined by $\abstr{u}_{n+1} = (\abstr{u}_n \widening \abstr{v}_n) \cap \abstr{K}$ is stationary if $\abstr{u}_n \subseteq \abstr{v}_n$ for all~$n$.
\end{definition}

This property ensures the correctness of widening ``up to'' \cite{Polka:FMSD:97}, a well-known improvement to widening, and is true of the standard widening on polyhedra as well as Bagnara et al.'s improved widening~\cite[p.~53]{BagnaraHRZ05SCP}. Using the same notations and hypotheses as above, we use this iteration:

\begin{equation}\label{eqn:restricted_iterations2}
\abstr{u}_{n+1} = (\abstr{u}_n \widening (\abstr{u}_n \sqcup (\abstr{\tfun}(\abstr{u}_n) \cap \abstr{K})))  \cap \abstr{K}
\end{equation}

Again, once we get a stationary value $\abstr{u}_N$ in this sequence, then it is such that $\lfp \tfun \subseteq \abstr{u}_N$:

\begin{lemma}
If $\abstr{u}_{N+1} \subseteq \abstr{u}_N$ in Eqn.~\ref{eqn:restricted_iterations2}, $\abstr{u}_N$ includes $\lfp \tfun$, the set of reachable states.
\end{lemma}

\begin{pf*}{Proof}
$\tfun(\abstr{u}_N) \cap \abstr{K} \subseteq
\abstr{\tfun}(\abstr{u}_N) \cap \abstr{K} \subseteq 
\abstr{u}_{N+1} \subseteq \abstr{u}_N$, from the correctness of $\abstr{\tfun}$.
Furthermore, by construction, $\abstr{u}_N \subseteq \abstr{K}$, thus
$\tfun(\abstr{u}_N) \cap \abstr{K} = \tfun_{|\abstr{K}}(\abstr{u}_N)$.
$\tfun_{|\abstr{K}}(\abstr{u}_N) \subseteq \abstr{u}_N$, thus $\lfp \tfun_{|\abstr{K}} \subseteq \abstr{u}_N$. The result follows from Lem.~\ref{lem:restriction}.
\end{pf*}

\section{Stratified Widenings} 
\label{sec:stratified_widening}
An alternative to the method described in the preceding section, which runs successive analyzes of increasing precision, is to run a single analysis over a \emph{reduced product} \cite{DBLP:journals/cl/CortesiZ11} of polyhedral domains, but with a special widening operator. We shall provide two options for that operator.

\subsection{Widening with or without Reduction}
\label{sec:stratified_widening_descr}
We distinguish the internal state $(P_S)_{S \in \calS}$ of the iteration sequence from the set of states represented, as in~\cite{Monniaux_HOSC09}. The various abstract operations will therefore continue operating on polyhedra as usual: only the widening operator is replaced.

Our widening operators will take a tuple $(P_S)_{S \in \calS}$ as a first argument and single polyhedron $Q$ as a second argument. A tuple $(P_S)_{S \in \calS}$ represents the polyhedron
\begin{equation}
\gamma\left((P_S)_{S \in \calS}\right) = \bigcap_{S \in \calS} P_S;
\end{equation}
the tuples are ordered point-wise, $(P_S)_{S \in \calS} \sqsubseteq (Q_S)_{S \in \calS}$ if and only if for all $S$, $(P_S) \subseteq (Q_S)$.

We note $\pi_S(P)$ the projection of polyhedron $P$ onto the variables in $S$. If $S \subseteq S'$, a polyhedron on the variables in $S$ shall be also considered as a polyhedron on the variables in $S'$ by keeping the same constraints. This means, in particular, that $P \subseteq \pi_S(P)$ for any $P$ and~$S$.

The first widening operator is very simple:
\begin{equation}
(P_S)_{S \in \calS} \widening_1 Q = (P_S \widening \pi_S(Q))_{S \in \calS}
\end{equation}
where $\widening$ is any widening on polyhedra. This widening converges because each coordinate converges, since $\widening$ is a widening. It is obvious that, if $(P_S)_{S \in \calS}$ is the resulting limit, then $\gamma\left((P_S)_{S \in \calS}\right)$ is an inductive invariant.

The second widening applies internal reductions. $(R_S)_{S \in \calS}$ denotes $(P_S)_{S \in \calS} \allowbreak \widening_2 \allowbreak (Q_S)_{S \in \calS}$. We compute the $R_S$ in ascending order with respect to $\prec$, with the convention that the intersection of zero polyhedra is the full polyhedron:
\begin{equation}\label{eqn:widening2}
R_S = (P_S \widening \pi_S(Q)) \cap \bigcap_{S' \prec S} R_{S'}
\end{equation}

\begin{theorem}
Assuming that $\widening$ is a widening satisfying the ``up to'' termination condition (Def.~\ref{def:upto-termination}), $\widening_2$ is a widening.
\end{theorem}

\begin{pf*}{Proof}
Let $u^{(n+1)}=u^{(n)} \widening_2 v^{(n)}$ be a sequence, with $u^{(n)} \sqsubseteq v^{(n)}$ for all $n$; each element $u^{(n)}$ consists in $u^{(n)}_S$ for $S \in \calS$. We prove that for all $S \in \calS$ the sequence $u^{(n)}_S$ is stationary, by induction over $\prec$.

For $S$ with no predecessor, $(u^{(n)}_S)$ is of the form $u^{(n+1)}_S = u^{(n)}_S \widening v^{(n)}_S$, and the result follows from $\widening$ being a widening.

Consider now the property satisfied for all $S' \prec S$. For all $S' \prec S$, $(u^{(n)}_{S'})$ is stationary; thus there is a $N$ such that for $n \geq N$, all $(u^{(n)}_{S'})$ for $S' \prec S$ are constant. $\bigcap_{S' \prec S} u^{(n)}_{S'}$ is thus constant for $n \geq N$. The results follows from $\widening$ being a widening satisfying our additional property.
\end{pf*}

Instead of polyhedra, one may use other abstract domains fitted with an operation $\sqcap$ such that $a \cap b \subseteq a \sqcap b$ for all $a,b$.
Let us however note that $\widening_1$ and $\widening_2$ yield the same results as the ordinary widening $\widening$ if applied to domains, such as difference bound matrices or octagons \cite{Mine_PhD} where $\widening$ and projection commute: $\pi_S(P) \widening \pi_S(Q) = \pi_S(P \widening Q)$, and therefore that they bring no improvement for such domains: the $P_S$ are just projections of $P_\calV$. More precisely:

\begin{lemma}
Assume that $\pi_S(P) \widening \pi_S(Q) = \pi_S(P \widening Q)$ for all $P$ and $Q$. Any iteration sequence of the form $P^{(n+1)} = P^{(n)} \widening Q^{(n)}$ then satisfies, for all $n$ and $S \in \calS$, $P^{(n)}_S = \pi_S(P^{(n)}_\calV)$, assuming this equality holds for $n=0$.
\end{lemma}

\begin{pf*}{Proof}
Regarding $\widening_1$: by induction over $n$, for any $S$, $P_S^{(n+1)} = P_S^{(n)} \widening \pi_S(Q^{(n)}) = \pi_S(P_\calV^{(n)}) \widening \pi_S(Q) = \pi_S(P_\calV^{(n)} \widening Q) = \pi_S(P_\calV^{(n+1)})$.

Regarding $\widening_2$: by induction over $n$, then by induction over $\calS$ with respect to $\succ$:
$(P^{(n)}_S \widening \pi_S(Q^{(n)})) \cap \bigcap_{S' \prec S} P^{(n+1)}_{S'} =
(\pi_S(P^{(n)}_\calV) \widening \pi_S(Q^{(n)})) \cap \bigcap_{S' \prec S} \pi_{s'}(P^{(n+1)}_\calV) =
\pi_S(P^{(n)}_\calV \widening \pi_S(Q^{(n)})) \cap \bigcap_{S' \prec S} \pi_{s'}(P^{(n+1)}_\calV) =
\pi_S(P^{(n+1)}_\calV) \cap \bigcap_{S' \prec S} \pi_{s'}(P^{(n+1)}_\calV) = \pi_S(P^{(n+1)}_\calV)$, since for any $X$ and $S' \succ S$, $\pi_{S'}(X) \cap \pi_S(X) = \pi_S(X)$.
\end{pf*}

\subsection{Generalized Reduction Leads to Nontermination}
Communicating information between several abstract domains used at the same time is sometimes referred to as a \emph{closure} or \emph{reduction} operation. Our $\widening_2$ operation includes a partial closure, with information flowing from $a$ to $b$ if $a \prec b$, but not the reverse. One could wonder about applying reductions in all directions. Unfortunately, we would lose the termination property of widening, as demonstrated by the following example.
\footnote{The fact that widenings followed by reductions with cycles (reduce $a$ using $b$, then reduce $b$ using $a$) may not ensure termination is already known. For instance, closure in difference-bound matrices and octagons breaks termination.~\cite[example 3.7.3, p.~85]{Mine_PhD}}

\begin{minipage}{8cm}
\begin{lstlisting}[label=lst:alterating_inc,caption={Alternating increments}]
int i=0, j=0;
while (true) {
  if (i <= j) i++; else j++;
}
\end{lstlisting}
\end{minipage}
\raisebox{-10mm}{\includegraphics[scale=0.9]{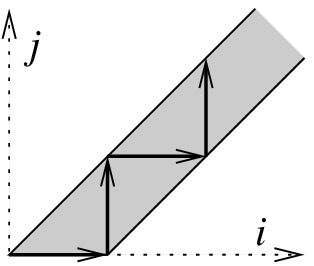}}

This loop has different behaviors on odd and even iterations: at iteration $2n$, $i=n$ and $j=n$; at iteration $2n+1$, $i=n+1$ and $j=n$. The results of a static analysis with polyhedra on $(i,j)$, and unions instead of widenings,
are, in constraint form: $\abstr{P}_{2n}: \abstr{P} \land i \leq n$ and $\abstr{P}_{2n+1}: \abstr{P} \land j \leq n$, $\abstr{P}$ denoting $i \geq j \land i \leq j+1 \land j \geq 0$ (we identify $\abstr{P}$ with the conjunctions of the constraints that define it). If for the iteration $n=4$ we use widening,%
\footnote{Applying unions at $n$ first iterations and then applying widening is a standard technique known as \emph{delayed widening}.}
we instead obtain $\abstr{P}_4 = \abstr{P}$, which is an inductive invariant.

We have established that this program poses no challenge to ``classical'' polyhedral analysis. The same is true if we apply one of the analyzes of Sec.~\ref{sec:stratified_analysis} or one of the widenings of Sec.~\ref{sec:stratified_widening_descr}. Let us now see what happens if we modify the $\widening_2$ operator of Sec.~\ref{sec:stratified_widening_descr} by allowing reductions not following~$\prec$.

Instead of the definition given at Eq.~\ref{eqn:widening2}, we instead initialize all $R_S$ to $P_S \widening \pi_S(Q)$, then apply some replacements, or \emph{reductions}, of the form:
\begin{equation}\label{eqn:widening2mod}
R_S := R_S \cap \bigcap_{S'\neq S} \pi_S(R_{S'})
\end{equation}
If we reach a fixed point for this replacement system, using the terminology from octagons \cite{Mine_PhD}, we say that we have applied the \emph{closure} operation.

Let us first remark that $\gamma\left((R_S)_{S \in \calS}\right)$ is left unchanged any number of such reductions:
\begin{lemma}
Let $(R'_S)_{S \in \calS}$ be the same as $(R_S)_{S \in \calS}$ except that $R'_{S_0} = R_{S_0} \cap \bigcap_{S'\neq S_0} \pi_{S_0}(R_{S'})$. Then, $\gamma\left((R'_S)_{S \in \calS}\right) = \gamma\left((R_S)_{S \in \calS}\right)$.
\end{lemma}

\begin{pf*}{Proof}
$\gamma\left((R'_S)_{S \in \calS}\right) = \bigcap_{S \in \calS} R'_S = \gamma\left((R_S)_{S \in \calS}\right) \cap \bigcap_{S'\neq S_0} \pi_{S_0}(R_{S'}) = \gamma\left((R_S)_{S \in \calS}\right) \cap \bigcap_{S'\in \calS} \pi_S(R_{S'}) $. Since $R_{S'} \subseteq \pi_{S_0}(R_{S'})$ for any $S'$, $\bigcap_{S'\in \calS} \pi_S(R_{S'}) \supseteq  \bigcap_{S'\in \calS} R_{S'} =  \gamma\left((R'_S)_{S \in \calS}\right)$. The result follows.
\end{pf*}

Because $\gamma\left((R_S)_{S \in \calS}\right)$ does not change, after the reductions, $\gamma\left((R_S)_{S \in \calS}\right)$ is still the same as $\gamma(P \widening Q)$. Our new ``widening'' thus verifies the soundness property (see Sec.~\ref{sec:formal}); the problem is that it does not verify the termination property!

Let us have $\calS= \{ \{i\}, \{j\}, \{i,j\}\}$; instead of $P_{\{i\}}$, $P_{\{j\}}$ and $P_{\{i,j\}}$ we shall respectively note $\abstr{I}$, $\abstr{J}$ and $\abstr{P}$. At iteration $n$, we shall therefore have a polyhedron $\abstr{I}_n$ on $\{ i \}$ (thus, an interval) and one polyhedron $\abstr{J}_n$ on $\{ j \}$ in addition to the polyhedron $\abstr{P}_n$ on $\{i,j\}$. If using unions instead of widenings, we have $\abstr{I}_{2n} = [0,n]$, $\abstr{I}_{2n+1} = [0,n+1]$, $\abstr{J}_{2n} = [0,n]$ and $\abstr{J}_{2n+1} = [0,n]$. Consider now using widening at the iteration $n=4$. $\abstr{I}_4 = \abstr{I}_3 = [0,2]$, but $\abstr{J}_4 = [0,+\infty)$. 

Let us now apply the closure operation: we replace $\abstr{P}_4 = \abstr{P}$ by  its intersection with $\abstr{I}_4$ and obtain $\abstr{P} \land i \leq 2$; then we replace $\abstr{J}_4$ by its intersection with the updated $\abstr{P}_4$ and obtain $[0,2]$. At the next iteration, with the roles of $\abstr{I}$ and $\abstr{J}$ reversed, we obtain $\abstr{I}_5=[0,3]$, $\abstr{J}_5=[0,2]$ after closure, and then $\abstr{I}_6=[0,3]$, $\abstr{J}_6=[0,3]$.

The iterations with widening followed by closure behave, on $\abstr{I}$ and $\abstr{J}$, like those with unions --- and \emph{they do not converge within finite time}.
Observe that this happens because we alternatively reduce $\abstr{I} \rightarrow \abstr{P} \rightarrow \abstr{J}$ and  $\abstr{J} \rightarrow \abstr{P} \rightarrow \abstr{I}$, whereas the definitions of Sec.~\ref{sec:stratified_widening_descr} only allow $\abstr{I} \rightarrow \abstr{P}$ and $\abstr{J} \rightarrow \abstr{P}$.

\section{Experimental Results} 

The stratified analysis presented in section~\ref{sec:stratified_analysis}, in
both variants (Eqn.~\ref{eqn:restricted_iterations} and
Eqn.~\ref{eqn:restricted_iterations2}), was evaluated against the classical
analysis described by Eqn.~\ref{eqn:iterations} on a set of benchmarks used by
STMicroelectronics in the development cycle of its compilers, in addition to a
few specific examples such as the one from Sec.~\ref{part:motivation_example}.

\emph{LAO Kernels} is a set of benchmarks internally used for the evaluation of
compilers code generators and optimizations. It is mainly composed of small
computational kernels representative of the target applications of
STMicroelectronics (audio and video stream processing, embedded device
control), associated with a testing harness to be able to run them on the
target processor. It contains 63 functions, of which 49 contain at least one
loop. Loops have to exhibit some properties, like a non-linear relation between
variables in the loop scope, in order to benefit from this method. Stratified
analysis finds a more precise invariant for 5 of these functions.

Among these 5 functions, \emph{discrete cosine transform}
has three nested loops. The intuition of why stratified analysis performs
better is it obtains an invariant for the indices affected by the outer loop
before attempting to analyze the inner loop, thus preventing imprecisions
during the inner loop analysis to affect the invariant on the outer loop
indices. 

The dependency relation used to create the strata is based on a modified dataflow graph; strongly connected components (SCC) are reduced to super-nodes, while keeping the existing dependency relations. Initial strata stem from the root nodes of this SCC dependency graph, additional ones are created by following the dependency relations until one stratum encompasses all variables in the dependency graph. In the \lstinline|while| loop of the listing~\ref{lst:loop42squared}, the variable \lstinline|j| depends from \lstinline|i|; the SCC nodes simply consist of $\left\{ i \right\}$ and $\left\{ j \right\}$, and the analysis creates two strata $\left\{ i \right\}$ and $\left\{ i, j \right\}$.

%
%

The two variants of stratified analysis described by
Eqn.~\ref{eqn:restricted_iterations} and Eqn.~\ref{eqn:restricted_iterations2}
find the same results, and in all cases find invariants equal to or stronger
than those obtained by the classical analysis.  Bagnara et al.'s alternate
widening \cite{BagnaraHRZ05SCP} yields iteration sequences different from those
obtained by the classical widening, but ultimately finds the same invariant;
thus, our approach improves on theirs on this benchmark set.

Table~\ref{fig:result_table} shows the number of variables in the
outermost stratum, along with the number of strata considered by the analysis
and its overhead with respect to the standard analysis using only the classic
widening. Some programs exhibit a large number of strata, impacting the cost of
the analysis. It is possible to run the expensive stratified analysis after a
first cheaper standard analysis, while focusing on certain loop nests (those
reaching $\top$ for instance).

\begin{table}[h]\footnotesize
    \begin{center}
        \begin{tabular}{|cccc|}
            \hline
            Function & $\#$ of vars & $\#$ of strata & Overhead\\ \hline
            autocorrelation & 9     & 8     & 5.55x\\ 
            binary search   & 2     & 2     & 1.95x\\ 
            discrete cosine transform             & 27    & 17    & 9.79x\\ 
            integer power   & 2     & 3     & 2.29x\\ 
            listing 2       & 2     & 2     & 1.66x\\ \hline
        \end{tabular}
    \end{center}
    \caption{Number of variable in the last stratum, number of strata and overhead of stratified analysis for programs that benefit from this method. The baseline for overhead measures is the classic analysis using bare widenings, without delay or widening-up-to).}
    \label{fig:result_table}
\end{table}



We rely on the \textsc{APRON} numerical abstract domain library%
\footnote{\url{http://apron.cri.ensmp.fr/library/}}
\cite{DBLP:conf/cav/JeannetM09} for all abstract domain computations. APRON
implements, among other domains, convex polyhedra with the classical widening,
with linearization of nonlinear expressions following Min\'e's approach
\cite{mine:vmcai06}. In addition, in order to compare with Bagnara et al.'s
alternate widening, we used the Parma Polyhedra Library%
\footnote{\url{http://www.cs.unipr.it/ppl/}} \cite{BagnaraHZ08SCP} (with the
classical widening, the PPL produces exactly the same results as APRON up to
equivalence of constraints, thus providing a means to test for possible bugs in
the polyhedral computations).

\section{Related Work} 
\label{sec:related_work}
It has long been recognized that analysis using polyhedra over all variables in a program, or even all variables in a single function, is unfeasible because of the high complexity of polyhedral operations in higher dimensions. This is also true of weaker domains such as octagons. For this reason, the Astr\'ee analyzer uses relational domains only on ``packs'' of variables~\cite{BlanchetCousotEtAl02-NJ,BlanchetCousotEtAl_PLDI03}: for instance, if we have four variables $a,b,c,d$ and two packs $\{a,b\}$ and $\{b,c,d\}$, the analysis will track relationships between $a,b$ and $b,c,d$ separately: no direct relation will be established between $a$ and $d$.

A related approach is \emph{factoring of polyhedra} \cite{DBLP:journals/fmsd/HalbwachsMG06}: when a polyhedron $P$ is a Cartesian product $P_1 \times \dots \times P_n$ of polyhedra in lower dimension, with respectively $v_i$ vertices (or, more generally, generators), it is often advantageous to keep this product representation as much as possible instead of considering it as a polyhedron of $\prod_i v_i$ vertices, because of algorithms that need to work on the generator representation. An alternative is to dispense totally with the generator representation \cite{DBLP:conf/sas/SimonK05,DBLP:conf/aplas/SimonC10}.

The literature on slicing is abundant, since the early 1980s~\cite{DBLP:journals/tse/Weiser84}. \emph{Syntactic slicing} extracts all program statements, variables etc. that affect the value of variable $v$, or, rather, a safe superset thereof. The resulting \emph{slice} is executable, which is interesting for testing or debugging methods, but less so for abstract interpretation; this is why we may use lax dependency relations (Sec.~\ref{part:strata}), since we in effect replace any unknown dependency by nondeterministic choice. \emph{Semantic slicing} relaxes the requirement that the resulting program be a syntactic subset of the original program~\cite{Ward:2007:SPT:1216374.1216375}. X.~Rival considers a form of abstract semantic slicing \cite{DBLP:conf/sas/Rival05,Rival_PhD}, where program executions are restricted to those affecting the reachability of undesirable program states (alarms); in contrast, our method does not suppose we have a set of properties (absence of alarms) to prove.

The design of widening operators is surprisingly difficult. The original widening operator on polyhedra \cite{CousotHalbwachs78} was sensitive to syntax: different ways of representing the same polyhedron in constraint form yielded different widened polyhedra; this problem was later fixed~\cite{Halbwachs_PhD}. Because the result of iterations with widening is non-monotonic, precision is highly heuristic: in particular, replacing a widening operator by one producing smaller polyhedra at each iteration does not necessarily translate in a smaller invariant in the end~\cite[p.~42]{BagnaraHRZ05SCP}.

Despite this caveat, many widening operators have been proposed for convex polyhedra~\cite[p.~30]{BagnaraHRZ05SCP}\cite{DBLP:conf/aplas/SimonC10}. Many are variants on the classical widening: some apply union in lieu of the classical widening in a way that does not preclude termination \cite{BagnaraHRZ05SCP}; the ``up to'' widening, also known as \emph{widening with thresholds} or \emph{limited widening} \cite{Polka:FMSD:97}, extracts possibly relevant constraints from the program and keeps in $P \widening Q$ the constraints from that set satisfied by both $P$ and $Q$; a related idea is \emph{widening with landmarks}, which uses estimates of the number of supplementary iterations necessary to enable a currently disabled transition \cite{Simon_King_APLAS06}; \emph{widening with a care set} uses a proof goal and counterexamples in order to guide the widening~\cite{Wang_et_al_CAV2007}. Our approach is largely orthogonal to these, and in fact can be combined with them.

In the recent years, there has been much interest in techniques for inferring invariants without doing conventional Kleene iterations. \emph{Policy iteration} (also called \emph{strategy iteration}; the technique is inspired by game theory) exists in two flavors. Descending policy iteration \cite{GGTZ:07} solves a descending sequence of least fixed points of simpler operators; these least fixed points may be solved approximately using widenings, thus this technique is orthogonal to ours. In contrast, ascending policy iteration \cite{Gawlitza_Seidl_ESOP07} and other techniques based on constraint programming \cite{Sankaranarayanan_PhD} or quantifier elimination \cite{Monniaux_LMCS10} provide some optimality guarantees, but impose restrictions on the kind of program instructions supported. Such restrictions may be lifted by abstracting program operations into the supported subset \cite{Mine_PhD}, which may in turn entail an outer loop with widenings.

We finally note that nothing in our approach is specific to polyhedra, or even to numerical domains. 

\section{Conclusion} 
Following our intuition that failure to analyze well parts of a program should not negatively influence precision on other parts not depending on them, we proposed four analysis schemes: two proceed by analyzes of restrictions of the program code to variable subsets, the other ones use alternative widening operators. Though we focused on improving the classical polyhedral analysis, two of our methods apply to any abstract domain, and the two other ones make a reasonable assumption on the underlying abstract domain and its widening operator.

\bibliographystyle{plainnat}
\bibliography{dependency_staged_analysis}

\end{document}